# Multi-Photon Transitions in Coupled Plasmon-Cyclotron Resonance Measured by Millimeter-Wave Reflection

Jie Zhang[1], Ruiyuan Liu[1,2], Rui-Rui Du[1,2], L. N. Pfeiffer[3], and K. W. West[3]

[1]*Department of Physics and Astronomy, Rice University, Houston, Texas 77251, USA*

[2]*International Center for Quantum Material, Peking University, Beijing 100871, China*

[3]*Department of Electrical Engineering, Princeton University, Princeton, New Jersey 08544, USA*

**Abstract**

We construct a low-temperature microwave waveguide interferometer for measuring high-frequency properties of two-dimensional electron gases (2DEGs). Coupled plasmon-cyclotron resonance (PCR) spectra are used to extract effective mass, bulk plasmon frequency, and carrier relaxation times. In contrast to traditional transmission spectroscopy, this method does not require sample preparation and is nondestructive. PCR signals can be resolved with a microwave power source as low as 10 nW. We observe PCR in the multi-photon transition regime, which has been proposed to be relevant to the microwave-induced resistance oscillations.

Microwave (MW) - induced resistance oscillation (MIRO) [1,2] and zero-resistance states (ZRS) [3,4] are spectacular phenomena observed in a non-equilibrium two-dimensional electron gas (2DEG) under millimeter-wave irradiation. Among possible mechanisms initially proposed to explain the origin of MIRO, *e.g.*, the displacement and the inelastic model [5,6], involve nonlinear effects. The observation that the MIROs correspond to the rational values of $\epsilon = \omega/\omega_c = 1/2, 1/3, 2/3 ...$ (where $\omega$ is the microwave frequency, $\omega_c = eB/m^*$ is the cyclotron frequency with $B$ the magnetic field, $m^*$ the effective mass of carriers) has been associated with multi-photon processes, especially at high MW intensities [7,8]. However, multi-photon cyclotron absorption associated with MIRO has so far not been clearly observed. It is of interest to study multi-photon process in the relevant regime, *i.e.*, with $\omega$ in the range of millimeter waves up to a few hundred GHz, and a small magnetic field $B < 10$ kG.

In the work presented here we observe coupled plasmon-cyclotron resonance (PCR) that involves multi-photon process, directly measured by MW reflections. We have measured a variety of electron and hole samples (see table I) in GaAs/AlGaAs heterostructures. Carrier effective mass, plasmon frequency and relaxation time are extracted from the data. Remarkably, we observed multi-photon absorption in all the electron samples, and also on the hole sample with higher density.

| Sample | Carrier type | Carrier density ($10^{11}/cm^2$) | Mobility ($10^6 cm/V \cdot s.$) | $(m*)$ | $\tau_t$(ps) | $\tau_s$(ps) | $\tau_t/\tau_s$ |
|---|---|---|---|---|---|---|---|
| **a** | Hole | 1.45 | 2.25 | 0.27 | 346 | 18 | 19 |
| **b** | Hole | 1.16 | 0.25 | 0.25 | 36 | 10 | 3.6 |
| **c** | Electron | 1.45 | 1.30 | 0.07 | 52 | ---- | ---- |
| **d** | Hole | 2.4 | 1.30 | 0.26 | ---- | ---- | ---- |

Table I   Parameters of hole samples **a**, **b** and electron sample **c**. Subscripts t and s refer to transport scattering time and single particle relaxation time, respectively. Line shape fitting for sample c and d is difficult because of the peak overlapping and shortage of samples.

Figure 1a. shows a schematic of the MW waveguide interferometer. The sample, which is typically a 3mm × 5mm square, is mounted to the end of a WG28 waveguide (TE$_{01}$ mode in the 26-40 GHz range) and backed with a copper plate, used due to its high reflectivity. The same waveguide is employed for both incident and reflected MWs. The MW source is a Model Anritsu 3694B. In order to separate two-way traffic, the collinear arms of a magic tee are adopted for MW input and output. Covering the remaining arm with an adjustable short allows for tuning the interference conditions. At $B = 0$ the two MW branches go through different paths, interfering at the output arm of the magic tee. In sweeping the magnetic field, the reflected MW from the sample is considered as a perturbation in an interference loop. In our experiments the MW source is modulated at 13 Hz, the Schottky diode detector signal is measured by a lock-in amplifier.

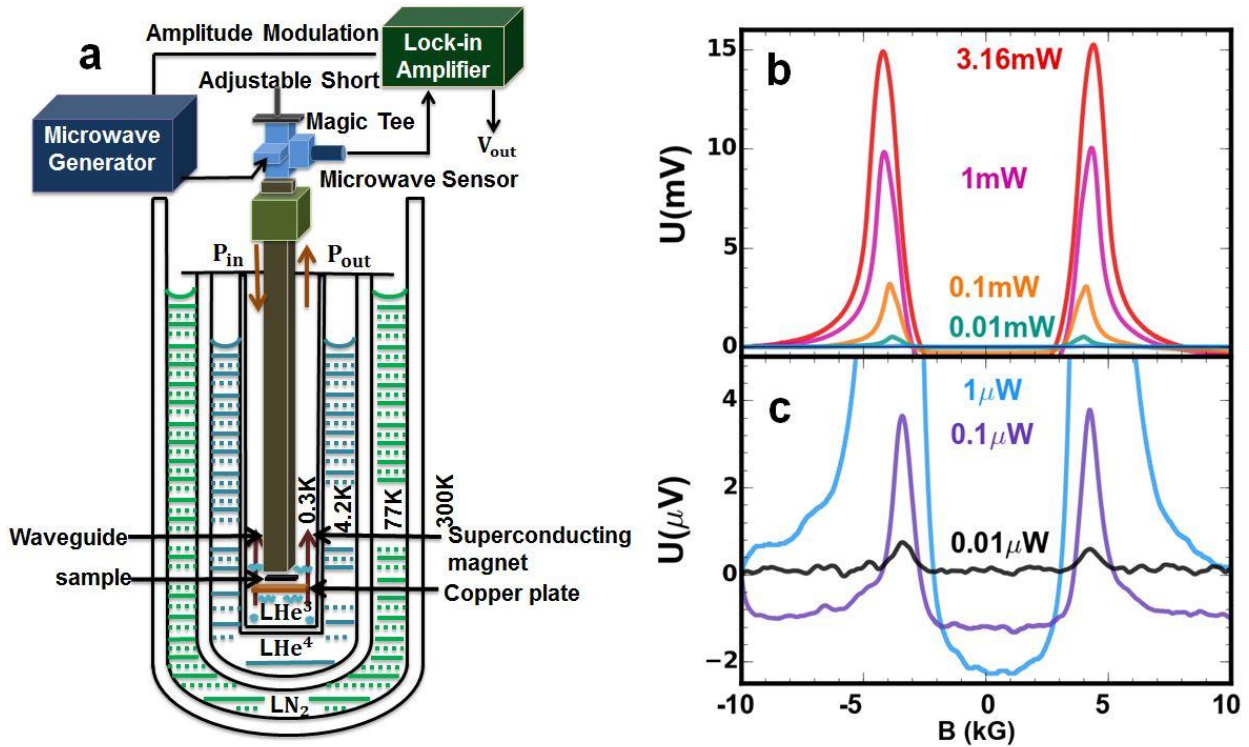

Fig.1 (a) a schematic of waveguide interferometer is shown. (b,c) Power dependence of PCR absorption on hole sample **a** with a fixed frequency of 40 GHz. For clarity, each data trace is shifted vertically by subtracting its value at -10 kG. The vertical axis is the amplitude of detector signal.

The probe which consists of the interferometer and the sample is top-loaded into a He3 cryostat with a base temperature of 300 mK; an axial magnetic field up to 8T can be applied by a superconducting magnet. The waveguide interferometer yields a clean background by partially cancelling the contribution of the source variation. Fig.1b shows that even with a source power of 10 nW (-50dbm), the cyclotron resonance signal still can be resolved, attesting to the high sensitivity and signal/noise ratio. Another advantage of the reflection method is that it obviates the need for sample preparation such as thinning/wedging the substrate, and hence preserves the quality of the wafer.

When a magnetic field is applied to the 2DEG, from Landau level (LL) spectrum

$$E = \hbar\omega_c(N_F + \frac{1}{2})$$

($N_F$ is the highest occupied level), a number of level transitions would be allowed near the Fermi surface. A sharp resonance occurs at $|N_F\rangle \to |N_F + 1\rangle$ where MW frequency matches $\omega_c$. Since the sample width is on the order of millimeter, contributions from bulk plasmons is comparable with energy level we are concerned and enter into the resonant frequency as coupled plasmon-cyclotron modes $\omega = \sqrt{\omega_c^2 + \omega_p^2}$. Here, the 2D bulk plasmon frequency is determined by the sample width $W$ via

$$\omega_p^2 = \frac{n_s e^2}{2m^* \epsilon_{eff}} \frac{N}{W},$$

where $N \in \mathbb{N}$ is the plasmon harmonic, $n_s$ is the sheet carrier density, and $\epsilon_{eff} = (\epsilon_0 + \epsilon_1)/2$ is the effective dielectric constant ($\epsilon_1$ is the dielectric constant of GaAs, $\epsilon_0$ is the dielectric constant of vacuum) [9]. Here either longitudinal resistance at integer filling factors or the slope of Hall resistance can be used to estimate carrier density (see Fig. 3).

Effective mass and bulk plasmon frequency are extracted by least squares fitting (Fig.2). Different-sized samples cut from the same wafer had identical effective masses but different plasmon frequencies. We found the effective masses of hole samples are ~ 0.25$m_e$ which are in the reasonable range between 0.2$m_e$ and 0.5$m_e$, depending on QW width, where $m_e$ is the free electron mass [10]. Electron samples generally have higher plasmon frequencies than hole samples due to their smaller effective mass (see Fig.4).

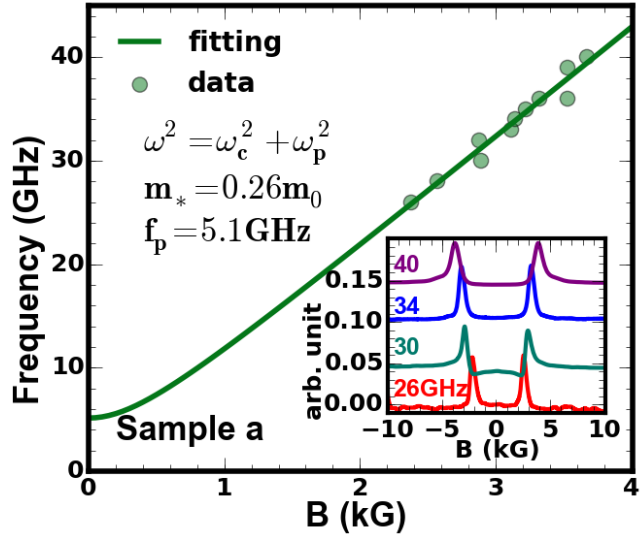

Figure 2 Frequency dependence of PCR absorption on hole sample **a** with a fixed power of 1 mW. Inset: For clarity, each data trace is first shifted vertically by subtracting its value at -10 kG, and then shifted upward consecutively by 0.025 unit. Effective mass and plasmon frequency are extracted using least square fitting.

Now we analyze the PCR line-shape; the sharpness of which can be quantified by its full width at half maximum (FWHM). This value corresponds to single particle relaxation time $\tau_s$ which is related to the imaginary part $\Gamma_s$ of the self-energy function [11] by $\Gamma_s = \hbar/2\tau_s$. Broadening of the LLs due to electron-impurity interaction in a Coulomb potential is responsible for broadening characterized by $\tau_s$. The transport scattering time $\tau_t$ from the semi-classical Drude model, on the other hand, is extracted from the DC conductivity using

$$\sigma_0 = n_s e^2 \tau_t / m^*$$

The cyclotron absorption coefficient $\alpha = \frac{1}{2} Re[\sigma_+(\omega) - \sigma_+(\omega)]$ is related to the dynamic conductivity via

$$\sigma_\pm(\omega) = \frac{i\sigma_0}{(\omega \mp \omega_c) + i/\tau_s}$$

where $\sigma_0$ is the DC conductivity in the Drude model. So the single particle relaxation time $\tau_s$ could be extracted from fitting the absorption trace with a Lorenztian function. It was shown that for a low mobility sample where scattering is mostly short-ranged and its cross-section is independent of scattering angle, these two time scales are equal [11]. But in the high mobility case, scattering is strongly peaked in the forward direction where transport scattering time $\tau_t$ can be one or two orders of magnitudes greater than the relaxation time $\tau_s$. A comparison between these two time scales is given in Table I with scattering time and relaxation time extracted from transport measurement (Fig.3 (a), (b)) and PCR Lorentzian fitting (Fig.3 (c)) respectively. We confirmed that $\tau_t/\tau_s$ increases with sample mobility. It should be emphasized that in order to use this method to extract relaxation time, we assume that small absorption approximation is valid. Cyclotron absorption will saturate when carrier density and mobility increases. In this case, the 2DEG behaves like a metallic sheet and reflects most of the incident microwave.

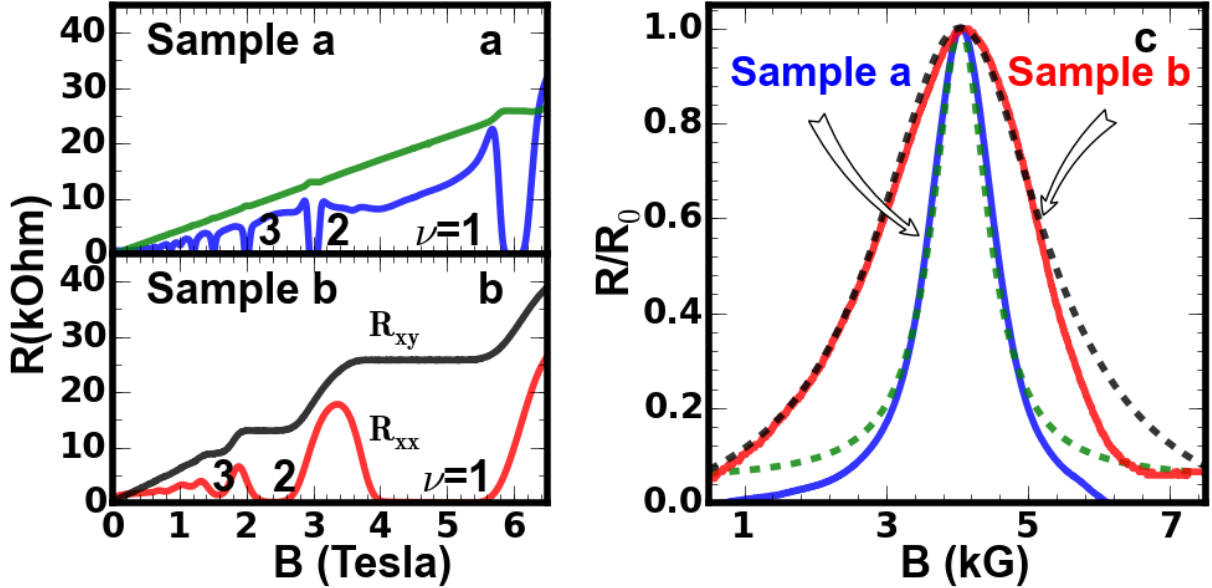

Figure.3 (a,b) Transport measurement of hole sample **a** and **b** for determining the transport scattering time $\tau_t$. (c) The data traces (solid lines) and Lorentzian fitting (dashed lines) of PCR line-shape to determine the single particle relaxation time $\tau_s$. FWHM of sample **a** and **b** are 0.94 kG and 2.89 kG, respectively. The data are taken with a MW frequency of 40 GHz at 1 mW. All time constants are presented in table I.

In the last part, we analyze multi-photon phenomenon we observed. This is a non-linear process characterized by higher order terms of the macroscopic polarization. High order contributions are usually several orders of magnitude weaker than the linear contribution but dominate at high intensities. It has been studied in photoresistance experiment recently by Dorozhkin *et al.* [12], and Zudov *et al.* [1] and are associated with fractional ratios

$$j \equiv \frac{\omega}{\omega_c} = n/m$$

where n and m are integers. We term the n > m case as high-harmonics while the case n < m sub-harmonics. In our case, when the sample has relatively high mobility and density, positions of PCR peaks split into values that could be interpreted as multi-photon process associated with sub-harmonic transitions where the resonant frequency is

$$\omega = \sqrt{(\frac{jeB}{m^*})^2 + \omega_p{}^2}, \text{(j=n/m)}$$

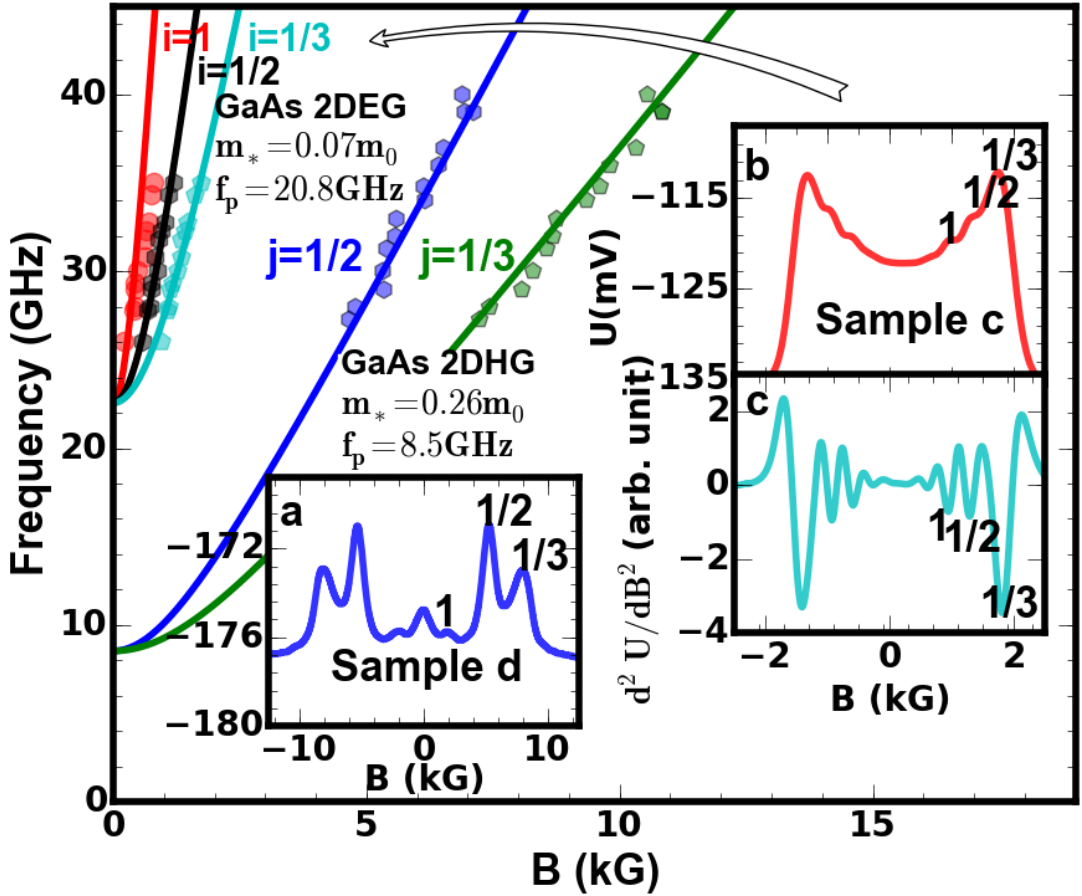

Figure.4 Multi-photon PCR on GaAs hole sample **d** and GaAs electron sample **c** data (scattered points) and fitting with dispersion relation (solid line). (a) Recorded data of hole sample **d** with MW of 31GHz and 1μW. Sub-harmonic orders are labeled; (b) Recorded data of electron sample **c** with WM of 34.8GHz and 0.1mW. Sub-harmonic orders are labeled. (c) 2$^{nd}$ derivative of data in (b).

For hole samples where the resonance peaks are widely separated, multiple-peak features are more visible (Fig. 4(a)). Electron samples however are generally more difficult to analyze since the peak separations are smaller than their FWHM (Fig.4 (b)) due to their smaller effective mass.

It can be seen from Fig.4(c) that by taking a second derivative of the recorded data to filter out the background contribution, the absorption peaks can be revealed.

We note that intensities of the sub-harmonic peaks generally do not show monotonic order, for example sample **d** in Fig. 4a. Phenomenologically, we found that multi-photon process is favored with increasing carrier density. (The 2DHG wafer used in Fig.4 has a hole carrier density of $2.4 \times 10^{11}/cm^2$.) In addition, we did not observe any high-harmonic peaks in the waveguide interferometer signal. This is consistent with the result of the non-interference method [13] and transmission measurement utilizing a bolometer.

In conclusion, we demonstrate that a low temperature waveguide interferometer is able to measure coupled plasmon-cyclotron modes in the gigahertz regime and small magnetic fields, including those involving multi-photon transitions. This method is an alternative to the standard transmission measurements in determining carrier effective mass and single particle relaxation time. Moreover, a combination of waveguide interferometer and transport experiment would open prospects for studies of microwave-induced resistance oscillations and the zero-resistance states.

*Acknowledgements* The work at Rice was supported by NSF Grant (No. DMR-1508644) and Welch Foundation Grant (No. C-1682). The work at Princeton was partially funded by the Gordon and Betty Moore Foundation as well as the National Science Foundation MRSEC Program through the Princeton Center for Complex Materials (No. DMR-0819860).